\documentclass[%
 reprint,
nofootinbib,
amsmath,amssymb,
aps,
prd,
superscriptaddress
]{revtex4-2}

\usepackage{graphicx}
\usepackage{dcolumn}
\usepackage{bm}
\usepackage{hyperref}
\usepackage{float}
\usepackage{subfig}
\usepackage{xcolor}
\usepackage{soul}
\usepackage{orcidlink}

\newcommand{\Ms}{M_\odot}

\begin{document}
\setlength\parindent{24pt}
\preprint{APS/123-QED}

\title{A neural network method to search for long transient gravitational waves}

\author{Francesca \surname{Attadio} \orcidlink{0009-0008-8916-1658}}
\affiliation{Universit\`a di Roma La Sapienza, 00185 Roma, Italy}
\affiliation{INFN, Sezione di Roma, 00185 Roma, Italy}

\author{Leonardo \surname{Ricca}}
\affiliation{Universit\`a di Roma La Sapienza, 00185 Roma, Italy}

\author{Marco Serra\orcidlink{0000-0002-6093-8063}} 
\affiliation{INFN, Sezione di Roma, 00185 Roma, Italy}

\author{Cristiano Palomba\orcidlink{0000-0002-4450-9883}} 
\affiliation{INFN, Sezione di Roma, 00185 Roma, Italy}

\author{Pia Astone\orcidlink{0000-0003-4981-4120}}
\affiliation{INFN, Sezione di Roma, 00185 Roma, Italy}

\author{Simone Dall'Osso\orcidlink{0000-0003-4366-8265}}
\affiliation{INFN, Sezione di Roma, 00185 Roma, Italy}
\affiliation{Universit\`a di Roma La Sapienza, 00185 Roma, Italy}

\author{Stefano Dal Pra\orcidlink{0000-0002-1057-2307}}
\affiliation{INFN, CNAF, 40127 Bologna, Italy}

\author{Sabrina D'Antonio\orcidlink{0000-0003-0898-6030}}
\affiliation{INFN, Sezione di Roma Tor Vergata, 00133 Roma, Italy}

\author{Matteo Di Giovanni\orcidlink{0000-0003-4049-8336}}
\affiliation{Universit\`a di Roma La Sapienza, 00185 Roma, Italy}
\affiliation{INFN, Sezione di Roma, 00185 Roma, Italy}

\author{Luca D'Onofrio\orcidlink{0000-0001-9546-5959}}
\affiliation{INFN, Sezione di Roma, 00185 Roma, Italy}

\author{Paola Leaci\orcidlink{0000-0002-3997-5046}}
\affiliation{Universit\`a di Roma La Sapienza, 00185 Roma, Italy}
\affiliation{INFN, Sezione di Roma, 00185 Roma, Italy}

\author{Federico Muciaccia\orcidlink{0000-0003-0850-2649}}
\affiliation{Universit\`a di Roma La Sapienza, 00185 Roma, Italy}
\affiliation{INFN, Sezione di Roma, 00185 Roma, Italy}

\author{Lorenzo Pierini\orcidlink{0000-0003-0945-2196}}
\affiliation{INFN, Sezione di Roma, 00185 Roma, Italy}

\author{Francesco Safai Tehrani\orcidlink{0000-0001-7796-0120}}
\affiliation{INFN, Sezione di Roma, 00185 Roma, Italy}




\begin{abstract}
We present a new method to search for long transient gravitational waves signals, like those expected from fast spinning newborn magnetars, in interferometric detector data. Standard search techniques are computationally unfeasible (matched filtering) or very demanding (sub-optimal semi-coherent methods).
We explored a different approach by means of machine learning paradigms, to define a fast and inexpensive procedure.
We used convolutional neural networks to develop a classifier that is able to discriminate between the presence or the absence of a signal.
To complement the classification and enhance its effectiveness, we also developed a denoiser.
We studied the performance of both networks with simulated colored noise, according to the design noise curve of LIGO interferometers.
We show that the combination of the two models is crucial to increase the chance of detection. 
Indeed, as we decreased the signal initial amplitude (from $10^{-22}$ down to $10^{-23}$) the classification task became more difficult. 
In particular, we could not correctly tag signals with an initial amplitude of $2 \times 10^{-23}$ without using the denoiser.
By studying the performance of the combined networks, we found a good compromise between the search false alarm rate (2$\%$) and efficiency (90$\%$) for a single interferometer.
In addition, we demonstrated that our method is robust with respect to changes in the power law describing the time evolution of the signal frequency.
Our results highlight the computationally low cost of this method to generate triggers for long transient signals.
The study carried out in this work lays the foundations for further improvements, with the purpose of developing a pipeline able to perform systematic searches of long transient signals.

\end{abstract}

\maketitle

\section{Introduction}
The first gravitational wave (GW) event, GW150914, was detected in 2015 \cite{bh_coalescence}, and was produced by the coalescence of two black holes with masses $\sim$ 29 $\mathrm{\Ms}$ and 36~$\mathrm{\Ms}$, respectively.
Since then, more than 90 events have been detected, all due to the inspiral and coalescence of neutron star (NS) or black hole binaries \cite{gwtc3}. 
Such signals are classified as short transients, with a duration lasting from a fraction of a second to a few tens of seconds in the band of LIGO \cite{LIGO}, Virgo \cite{VIRGO} and KAGRA \cite{KAGRA} ground-based detectors.
\par
Several other source classes are searched \textcolor{black}{for} in GW data \cite{gw_source}, and have not been detected yet.
Among them, there are sources of continuous quasi-periodic GWs (CW), e.g., asymmetric spinning neutron stars \cite{cw_ref}. 
For such sources, the signal frequency is linked to the star's rotational frequency by a proportionality factor and slowly decreases in time at a rate (i.e. spin-down) $ \lesssim (10^{-10}\mathrm{Hz}/\mathrm{s})$, due to the emission of GW itself and likely other energy losses, like the emission of electromagnetic (EM) radiation.
\textcolor{black}{So far, interesting upper limits on the amplitude of possible CW signals have been placed} (see e.g., \cite{Dergachev:2024knd,DOnofrio:2023amp,LIGOScientific:2022enz,LIGOScientific:2021quq,upper_limit_cw1, upper_limit_cw2,Steltner:2023cfk,Dergachev:2022lnt,LIGOScientific:2021inr,LIGOScientific:2021hvc,LIGOScientific:2021ozr,LIGOScientific:2021yby}).
\par
A variety of models (e.g. \cite{palomba2001,cutler2002, Dall_Osso_2021,askell2022}) have been proposed for millisecond spinning, \textcolor{black}{highly asymmetric NSs} (likely, newborn magnetars) emitting quasi-periodic GW signals over a limited time interval (in the detector's band), from minutes to several hours, with a spin-down $\sim \mathcal{O}(\mathrm{Hz}/\mathrm{s})$. 
Such signals are generically called \textit{long transients} \textcolor{black}{(or transient continuous waves, tCWs)}\cite{long_tra_Riles_2023}.
As we currently do not know exactly how matter behaves at supra-nuclear densities, the detection of this kind of signals will be a powerful additional tool in the study of the NS interior structure and behavior.
Typical search techniques used for tCWs, based on semi-coherent methods (see e.g., \cite{gfh_method, STAMP, hmm,adaptive_fh}, applied for \textcolor{black}{instance} in \cite{post_mergergw170817, gw170817}) or on matched filter (see e.g., \cite{matched_filter1_keitel, matched_filter2_keitel}), are computationally demanding.
\par
Here we propose a machine learning (ML) technique to perform a robust and computationally feasible search and produce a first list of candidates.
With the exception of the initial studies done in \cite{miller,keitel_glitch} there are no ML procedures implemented for this kind of signal.
In general, data analysis using neural networks (NNs) requires less computing power than more classical algorithms. 
Indeed, once a model\footnote{Model aka neural network in this paper.} is trained, it can be applied to different cases, as we will see in the subsection \ref{subsec:braking_index}.
\par
The aim of this work is to develop ML techniques to analyze ground-based interferometric detector data. 
This paper presents the implementation of a NN \textit{classifier} capable of identifying
the presence or absence of tCW signals.
Data are represented using time-frequency maps, \textcolor{black}{inputs to} the classifier model built with convolutional neural networks (CNN) \cite{cnn2020survey}. 
An additional neural network model has been placed before the classifier itself to improve its performance, specifically a NN \textit{denoiser} model exploiting residual learning \cite{res_learning1}.
\par

The paper is organized as follows. 
In \textcolor{black}{Sec.} \ref{sec:signal}, we describe the main features of the tCW expected from a newly born magnetar.
In \textcolor{black}{Sec.} \ref{sec:methods} we explain how we build time-frequency maps, input to our artificial neural networks, and how we choose the parameter space to simulate the signal.
In \textcolor{black}{Sec.} \ref{sec:results}, we report the results of our study, describing first the denoiser and classifier performances and subsequently the robustness of our models. 
Finally, we draw conclusions in \textcolor{black}{Sec.} \ref{sec:conclusions}.\par

\section{Signal model} \label{sec:signal}
The rotational energy of an isolated NS is dissipated through the emission of EM and/or gravitational radiation, implying a time variation of the rotational frequency described by the spin-down equation 
(e.g., \cite{hamil2015braking_index}): 
\begin{equation}
    \dot{f}_{\rm rot}=-k f_{\rm rot}^{n}
    \label{Eq:spin_down_general}
\end{equation}
where the spin-down constant $k$ and the braking index $n$ are both determined by the emission mechanism.
The value of the braking index is related to the emission process, and the constant contains information about the cause of the emission. 
For example, in the case of magnetic dipole emission, $n=3$, and $k$ depends on the strength of the dipolar magnetic field (and on the NS moment of inertia, $I$).
\par
In general, an isolated NS spinning with frequency $f_{\rm rot}$ can emit GWs if it presents an asymmetry with respect to the spin axis.
GW emission due to the NS time-varying mass quadrupole corresponds to  $n = 5$ (e.g., \cite{braking_indexMagnetars}), \textcolor{black}{although some instabilities, e.g.} $r-$modes, 
\textcolor{black}{lead to GW emission with} $n=7$.
If, in addition, the rotational energy is radiated only via GWs, then the spin-down equation (\ref{Eq:spin_down_general}) becomes 
\begin{equation}
    \dot{f}_{\rm rot}=- k_{GW} f^5_{\rm rot},      \ \ \ \ \ \ k_{GW}=\frac{32}{5} \frac{G I \epsilon^2 \pi^4}{c^5}
    \label{eq:spindown}
\end{equation}
where $G$ is Newton's gravitational constant, $c$ is the speed of light, $I$ is the moment of inertia along the rotational axis, and $\epsilon$ is the ellipticity, a measure of the star \textcolor{black}{asymmetry defined as}
 \begin{equation}
     \epsilon \equiv \frac{|I_1-I_2|}{I},
 \end{equation}
 where $I_1$ and $I_2$ are the moments of inertia around the other two principal axes.
 \par
 Magnetars are a particular class of NSs, characterized by a strong external magnetic field, $10^{14}-10^{15}$ G, and an even stronger interior component\cite{Dall_Osso_2021}.
They originate from a (significant) fraction, at least $\sim 10$\% \cite{mag_population} of stellar collapses, and are also expected to be formed as remnants in (a minority of) binary NS mergers, depending on details of the equation of state (EoS) of NS matter \cite{mag_from_ns,mag_from_ns2, mag_formation,Lasky_2019_nag_from_ns}. 
Their ellipticity, at formation, is expected to be $\sim 10^{-4}-10^{-3}$, the order of the magnetic-to-binding energy ratio (e.g. \cite{braking_indexMagnetars}), when the super-strong magnetic field is the main source of anisotropic stresses \footnote{Hydrodynamical instabilities like, e.g., the secular bar-mode instability may lead to an even larger degree of asymmetry, although shorter-lived (\cite{CorsiMezaros2009} and references therein)}.\par
The solution of Eq. (\ref{eq:spindown}) is
\begin{equation}
    f_{\rm rot}(t)= f_{0, {\rm rot}} \bigg(1+ \displaystyle \frac{t}{\tau}\bigg)^{-\frac{1}{4}} \ \ \ \ \ \tau=\displaystyle \frac{1}{4 f_{0, {\rm rot}}^{4} k_{GW}} \, ,
    \label{eq:freq_variation}
\end{equation}
where $f_{0, {\rm rot}}$ is the initial frequency and $\tau = f_{0, {\rm rot}}/4\dot{f}_{0, {\rm rot}}$ the characteristic spin-down time.\par

The source was modelled as an ellipsoid that rotates along one of its principal axes
for which the GW frequency $f(t)$ is linked to $f(t)_{\rm rot}$ by \cite{Maggiore}
\begin{equation}
    f(t)=2 f(t)_{\rm rot} .
\end{equation}
The GW signal received by an observer at a distance $d$, whose line of sight makes an angle $\iota$ with the direction of the spin of the star, is described in terms of the two wave polarization modes:
\begin{equation}
\begin{aligned}
&h_+(t)=h_0(t) \frac{1+\cos^2 \iota}{2} \cos(2 \pi t f(t) )\\    
&h_x(t)=h_0(t) \cos \iota \ \sin(2 \pi  t f(t)) \, ,
\end{aligned}
\label{eq:wave_eq_polarized_ellissoid1}
\end{equation}
where
\begin{equation}
\label{eq:wave_amplitude_ellipsoid}
    h_0(t)=\frac{4 \pi ^2 G}{c^4} \frac{I f(t)^2}{d} \epsilon.
\end{equation}

\section{\label{sec:methods}Methods} 
The \textcolor{black}{method we present involves the implementation of an NN that properly initialized allows the presence of a tGW to be distiguished from the noise present in the data produced by an interferometer. The data format we use and the NNs implemented are described here.}

\subsection{Introductory concepts of machine learning}
\textcolor{black}{Machine learning is a type of Artificial Intelligence (AI) that allows computers to learn without being explicitly programmed \cite{ref_giagu}. It involves feeding data into algorithms (implemented in a network \textit{model}) that can then identify patterns and make predictions on new data. ML implementations are classified into four major categories, depending on the nature of the learning “signal” or “response” available to a learning system. Among these we use
\textit{Supervised Learning}, that is the ML task of learning a function that maps an input to an output based on example input-output pairs. 
Both "classification" and "denoising" problems are supervised learning problems, solved with specific models (\textit{classifiers, denoisers})
A \textit{classifier} is a model that tries to predict the correct category between distinct classes(\textit{"labels"}) of a given input data. 
The \textit{denoiser} task involves removing noise from an image (from a data representation in which signal and noise are present). In our task, the models are fully initialized to use (\textit{"trained"}) using the labeled training data(\textit{signal vs absence of signal}).
Before we get to the implementation of our models we introduce the data format we use in the next paragraph.
}

\subsection{Time-frequency maps}\label{subsec:time_freq_maps}
A time-frequency representation of data is a well suited starting point for the search of tCW signals.~We have developed a MATLAB code to generate various kinds of time-frequency maps, in particular spectrograms, exploiting the \textcolor{black}{SNAG} package \cite{Snag}.~The code starts from the data time series, divides it into multiple segments, and then, for each chunk, using the Fast Fourier Transform (FFT) algorithm, computes an estimator of the power spectrum, namely the periodogram, which is the square modulus of the Fourier transform.
\begin{figure*}
\includegraphics[scale=0.37]{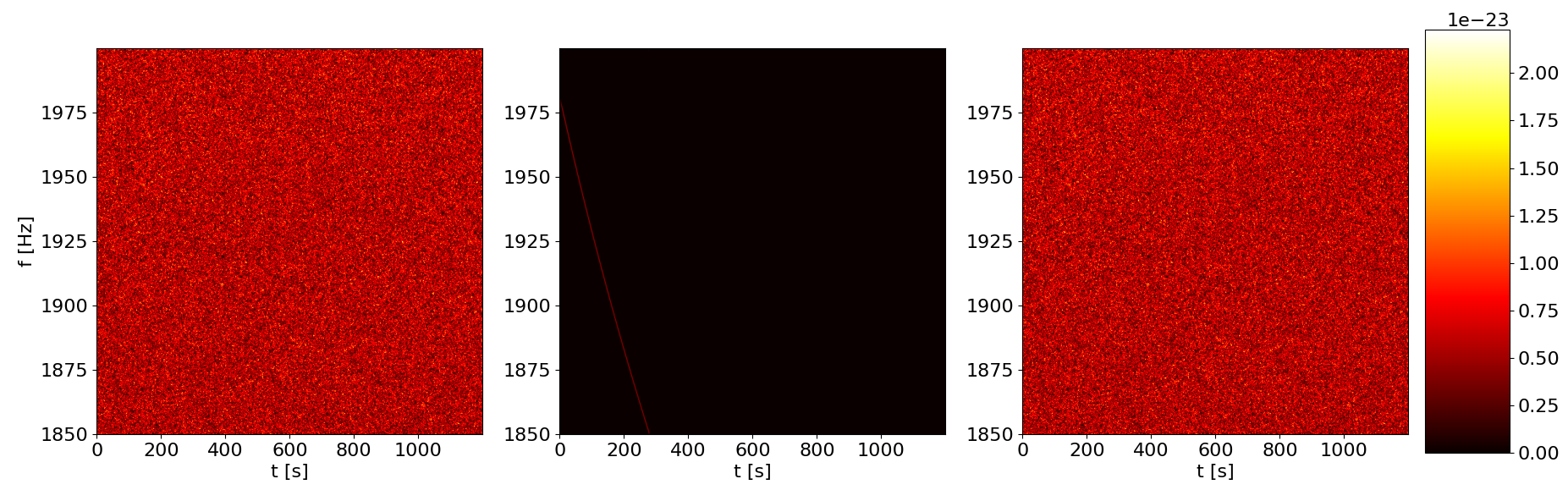}
\caption{\textcolor{black}{Example} of spectrograms of noise plus signal, only signal and only noise. The physical parameters for the signal are $f_0$=1982 Hz, $\epsilon$=0.0028, $h_0=2\times 10^{-23}$. Each map has a frequency range between 1850 Hz and 2000 Hz and covers a time interval of 1200 s. The noise is simulated according to the O4 \textcolor{black}{(begin: May 24, 2023, end: planned June 2025)} design sensitivity curve of LIGO interferometers.}\label{fig:spec_trittico}
\end{figure*}

\textcolor{black}{Figure} \ref{fig:spec_trittico} shows examples of spectrograms, with time~on the horizontal axis and frequency on the vertical axis, computed from simulated colored noise  \cite{colored_noise} following the LIGO O4 design sensitivity curve \cite{O4_sensitivity}.
The color bar corresponds to amplitude spectral density (ASD), i.e. the square root of the spectrum, associated with each pair of time and frequency values.\par
The spectrogram can be represented as a matrix where each column corresponds to the modulus of the FFT of a time segment of duration $\Delta t$.
The size of the frequency bin is
\begin{equation}
    \delta f =\frac{1}{ \Delta t}.
    \label{eq:freq_bin}
\end{equation}
Given that we are computing our FFT interlaced by half, the number of FFTs in a time interval T is
\begin{equation}
    N_t=\mathrm{floor}\left(\frac{2T}{ \Delta t}\right)
\end{equation}
and, hence, our time bin $\delta t$, i.e. our time resolution, is
\begin{equation}
    \delta t =\frac{ \Delta t}{2}.
\end{equation}
The choice of the time bin $\delta t$ is based on the characteristics of the signals. 
Ideally, in order to maximize the signal-to-noise ratio, one would choose the longest duration, which still allows the signal power to remain confined in a single frequency bin, i.e. ($\dot{f} \delta t  \leq \delta f =  1/(2\delta t)$). 
However, different signals can have widely different values of $\dot{f}$, i.e. different slopes in the time-frequency plane, depending on the signal parameters.
Given the parameter space we explore (see Eq. (\ref{eq:real_source_intervals}) and related discussion), after various tests, we have concluded that a reasonable choice is to use a segment duration $\delta t$ = 2 seconds.
In addition, in order to build squared maps, we computed each spectrogram over a time interval of 1200 s and a frequency band of 150 Hz.
In this way, we have maps of 600$\times$600 pixels, where each pixel has a single value assigned.
So a time-frequency map can be represented as a \textcolor{black}{gray-scale} image.
The choice of this representation was suggested by the properties of the signal we are searching for, as discussed later in this section.
The size (600$\times$600) was also chosen also taking into account the maximum memory that can be used on the GPU in NN training and the layers we used to implement the model.\par
Since we are computing the FFT of segments of finite duration, we have to account for spectral leakage, \textcolor{black}{which consists of a spread} of the power content of a frequency bin among neighboring bins.
To reduce spectral leakage, \textcolor{black}{we adopted a flat-top window}, defined as a sum of cosine functions:
\begin{equation}
\begin{aligned}
    & f_{\rm cos}(x_i) = \frac{1}{2} \bigg(1- \cos \bigg(\frac{4 \pi x_i}{N} \bigg) \bigg) \ \ \ \ i \leq \frac{N}{4}, \ i \geq \frac{N}{4}\\
    & f_{\rm cos}(x_i) =1 \ \ \ \ \ \ \ \  \ \ \ \ \ \ \ \ \ \  \ \ \ \ \ \ \ \ \ \ \ \ \ \ \ \ \frac{N}{4}<i<\frac{N}{4}\\
\end{aligned}
\end{equation}
where $N$ is the number of samples in the segment duration considered and $x_i$ are the samples ($i=1,..,N$). \par
To train the NN needed for the analysis, we created three types of maps containing, respectively, only noise, only the signal and the sum of noise plus signal, as shown in \textcolor{black}{Fig.} \ref{fig:spec_trittico}, from left to right.~To 
simulate the signal, according to Eq. (\ref{eq:wave_eq_polarized_ellissoid1}), we fixed a fiducial value for the NS \textcolor{black}{momentum of inertia}, 
$I= 1.4 \times 10^{38} \ \mathrm{kg \  m}^2$.
As for the angle of sight, we used $\iota=55^{\circ}$, \textcolor{black}{representative of the average inclination.}
Finally, we set the following ranges for the initial frequency ($f_{0}$) and NS $\epsilon$, respectively:
\begin{equation}
     f_0 \in [1.25,2.00] \ \mathrm{kHz}\ \ \ \ \  \epsilon \in [3,30] \times 10^{-4} \, ,
     \label{eq:real_source_intervals}
\end{equation}
which resulted from a combination of computational considerations, memory usage, signal shape and astrophysical arguments (see, e.g., \cite{mag_simone_cristiano}).

\begin{figure*}[hbtp]
    \centering
    \subfloat{\includegraphics[scale=0.37]{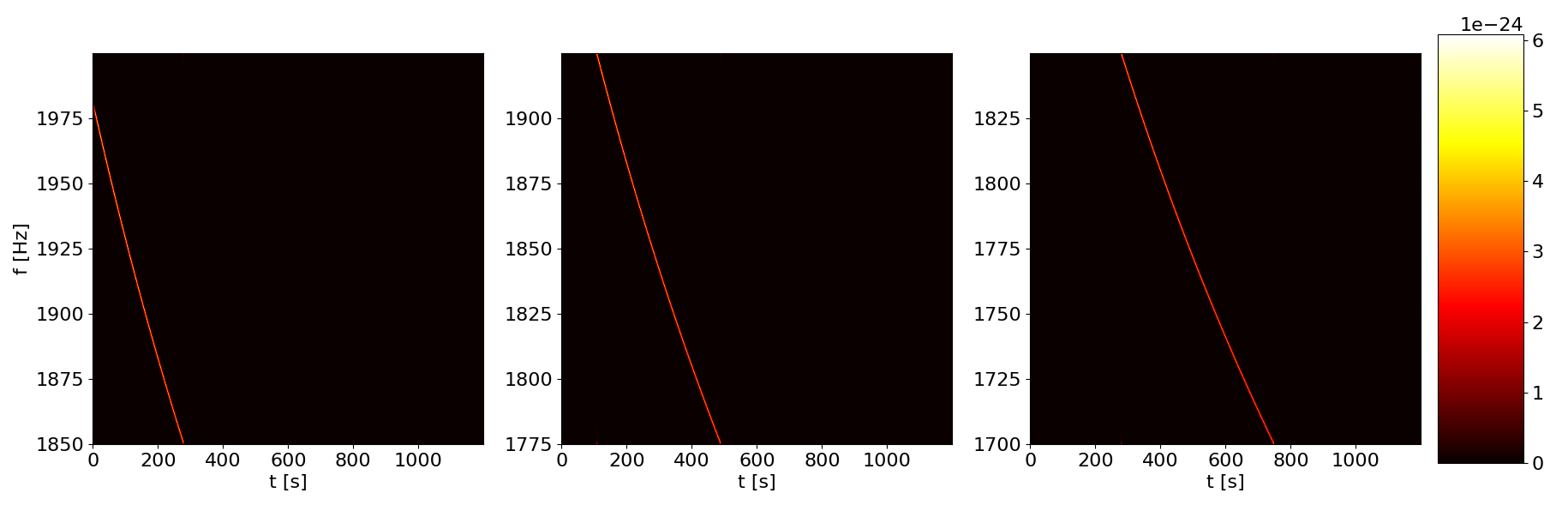}} \\
\subfloat{\includegraphics[scale=0.37]{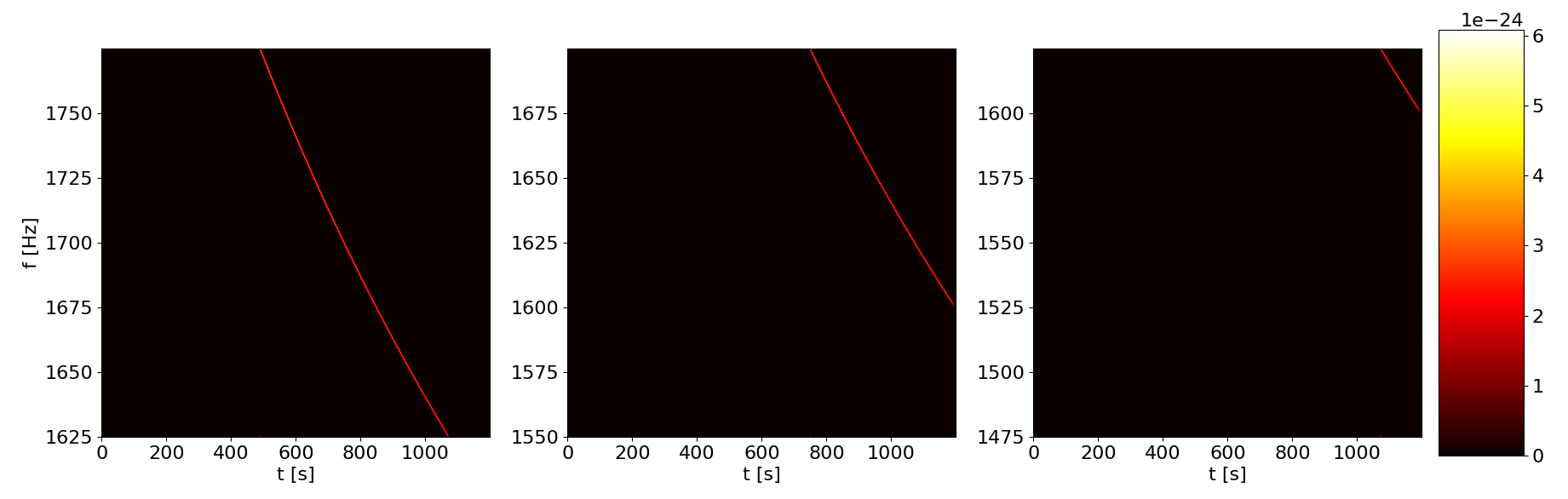}}\par
    \caption{\textcolor{black}{Signal simulated with} $f_0=1982$ Hz and $\epsilon=0.0028$. Different frequency intervals, same time interval. The first, third and fifth maps cover the frequency range from 2075 Hz to 1625 Hz, while the second, fourth and sixth maps cover the frequency range from 2000 Hz to 1550 Hz. We count the maps starting with the one in the upper left angle and we move subsequently to the right.}
   \label{fig:signal}
\end{figure*}

As shown in Eq. (\ref{eq:freq_variation}), different values of $f_0$ and $\epsilon$ lead to widely different signal shapes in the time-frequency plane.
In particular, a higher initial frequency and/or larger ellipticity imply a much faster frequency evolution, which, in some cases, may decrease by over 150 Hz in less than $1200$ s. 
Such signals would therefore cross multiple maps. 
\textcolor{black}{To} augment the chances of retrieving them, maps were interlaced by 75 Hz. 
As an example, \textcolor{black}{Fig.} \ref{fig:signal} depicts a signal with $f_0=1985\ \mathrm{Hz}$ and $\epsilon=2.8 \times 10^{-3}$ covering a range wider than 300 Hz in 1200 s, thus appearing in 6 interlaced maps.\par

The last degree of freedom is the distance to the magnetar or, equivalently, the initial GW amplitude.
We simulated signals at different fixed initial amplitudes, and here we report only the case with $2\times10^{-23}$, our current limit, which is the minimum amplitude for which the use of the denoiser turned out to be beneficial to train the classifier.
At lower amplitudes (using the O4 noise curve), our NNs are not able to discriminate between noise and signal.\par
To quantify the signal strength relative to noise,  we compute the pixel-signal-to-noise ratio ($pr_{\rm sn}$) as
\begin{equation}
    pr_{\rm sn}=\frac{1}{N_{\rm pix}} \sum_{N_{tf}\neq 0} \frac{S_{tf}}{N_{tf}},
    \label{eq:prsn}
\end{equation}
where $N_{\rm pix}$ is the number of pixels in the map, $N_{tf}$ is the noise amplitude spectral density 
at each map pixel, with time $t$ and frequency $f$, and $S_{tf}$ is the signal amplitude spectral density at the same map pixel.

\subsection{Machine learning models}
ML provides different approaches to solving complex problems in numerous areas \cite{ref_giagu}. 
The characteristics of the data and the specific problem define the best class of models for accomplishing the desired task.
Deep learning (DL) is a specific technique to implement ML models that focuses on training multi-layered artificial neural networks (ANNs) to learn complex patterns and representations from data.
DL is particularly well suited~to~tasks involving large amounts of data, such as “denoising" and “image classification", because of its ability to automatically discover nonlinear patterns.
In particular, we exploited CNNs \cite{CNN} to build our deep NNs, specifically a signal denoiser and a classifier.
The layer structure of the CNN is not densely connected, i.e., not all input nodes affect all output nodes.
As a result, the number of parameters (weights) needed to characterize a layer is smaller than in a typical dense architecture, \cite{bishopdeeplearning}, which helps with high-dimensionality inputs such as the image data we use to represent the signal. 
For this reason, CNNs require less computational power to learn the characteristics of the data than other artificial neural networks. 
To implement our NN models, we used the \textcolor{black}{Pytorch} framework \cite{pytorch}.

\subsubsection{Denoiser}\label{subsec:denoiser}

The goal of a denoiser, a specialized ANN, is to retrieve a clean image $y$ from a noisy observation $x=y+\nu$, where $\nu$ is the noise.~We adopted the residual learning approach (see \cite{Zhang_2017} and \textcolor{black}{Appendix} \ref{denoiserArch} for details on our modified architecture).
The model is closed by a skip \textcolor{black}{connection}\footnote{Technique used to link network input data directly to the last stages by skipping some branches (\cite{skip_connection}).} that links the output to the input.
The idea is to teach the model how to predict noise and then subtract it from noisy data images in order to produce a denoised map, i.e. the input for the classifier. \par

A supervised training approach was followed.
As a loss function, i.e., a mapping from real events into a real number, we choose the mean square error:
\begin{equation}
    MSE=\frac{1}{N_{\rm samples}} \sum_{i=1}^N (\hat{y}_i-y_i)^2,
    \label{eq:mse_loss}
\end{equation}
where $N_{\rm samples}$ is the number of samples, $y_i$ is the predicted value and $\hat{y}_i$ the ground truth.
Our goal is to minimize this quantity to find the best set of parameters.\par
An optimal denoiser should be able to reduce the noise level of the image without degrading the signal.
To evaluate its ability to preserve the signal, we extended to two dimensions the waveform overlap used in \cite{overlap}:
\begin{equation}
    \mathcal{O}=\sqrt{\sum_{tf} h_{tf}h_{tf}^d \bigg( \sum_{tf} h_{tf}h_{tf} \bigg)^{-1}}
    \label{eq:over}
\end{equation}
where $h_{tf}h_{tf}^d$ is the multiplication pixel by pixel of the signal map (the ground truth, $h_{tf}$) and the output of the denoiser ($h_{tf}^d$) while $h_{tf}h_{tf}$ is the multiplication pixel per pixel of the signal map with itself.
The overlap values range from 0, when no track of the signal has been preserved, to 1, when the entire signal has been retrieved.

\subsubsection{Classifier}\label{subsec:class}
An image classifier is a model that can distinguish different images that have different labels \cite{imageclassifier}.
In particular, we are interested in a binary classifier: the two classes are the presence of a signal (positive class) and the absence of a signal (negative class). 
So to train our model, we used the binary cross entropy as a loss function (see \cite{binary_cross_entropy} for more details).

The receptive field in the first layer, as shown in \textcolor{black}{Appendix} \ref{classifierArch}, is bigger for the classifier than for the denoiser because we want to identify a correlation between distant pixels on the scale of the whole map.
The signal differs from the noise because of this correlation, so this is a key feature that we want to extract to classify the maps.\par
The output of the NN is interpreted as the probability of the presence/absence of a signal. 
The samples that have been accurately predicted are either true positives (TP) or true negatives (TN). 
On the contrary, mismatched samples are known as false positives (FP) and false negatives (FN).\par
We studied the performance of the classifier, in particular the efficiency of detection (Eff) and the false alarm probability (FAP), adopting different probability thresholds to discriminate between the presence or absence of a signal in a given map. 
In principle,  \textcolor{black}{the Eff should be computed as}
\begin{equation}
    Eff=\frac{TP}{TP+FN}.
\end{equation}
Nevertheless, as discussed in subsection \ref{subsec:time_freq_maps}, a signal can cross multiple maps when its frequency varies very rapidly. 
\textcolor{black}{We define a "positive trigger" to select a GW candidate when at least one map is correctly tagged.}
So, from here on, we will report the efficiency per signal (Eff), defined as the number of positive triggers over the total number of injected signals. 
For what it concerns the FAP, we computed it as
\begin{equation}
    FAP=\frac{FP}{FP+TN+TP+FN},
\end{equation}
where the denominator is the sum of all the samples considered.
Eventually, to obtain a good balance between Eff and FAP, we used the F1 score, \cite{f1score}, defined as:
\begin{equation}
    F_1=\frac{2}{\frac{1}{P}+\frac{1}{R}}=2 \times \frac{P \times R}{P+R}=\frac{TP}{TP+\frac{FN+FP}{2}}.
    \label{eq:f1score}
\end{equation}

where the \textit{precision} P and the \textit{recall} R are defined as

\begin{equation}
    P=\frac{TP}{TP+FP} \ \ \ \ \ \ R=\frac{TP}{TP+FN} \, .
\end{equation}

\textcolor{black}{F1 is a harmonic mean, so it gives more weight to lower values of P and R, and it is large only if both P and R are large.}
In particular, it is 0 if there are no true positives, i.e. Eff $\sim$ 0 and FAP $\sim$ 1, and 1 if there are no false predictions, i.e. Eff $\sim$ 1 and FAP $\sim$ 0.
So we need to maximize the F1 score.\par
We use ROC curves to provide a visual representation of our classifier's performance. 

\section{Results}\label{sec:results}
\subsection{Dataset preparation}
In order to train the two models, we built two sets of maps.
Given that we were using two NNs at a time, the testing set of the first one (the denoiser) has been split to become the training and testing set of the second one (the classifier). 

To train the denoiser, we simulated 1000 signals in the range described by Eq. (\ref{eq:real_source_intervals}) and 200 more with initial frequency $f_0$ in the frequency interval [1.8, 2.0] kHz.
To test the denoiser and later train and test the classifier, we built another set of maps, with 2400 signals in the same intervals as Eq. (\ref{eq:real_source_intervals}).
We trained and tested our models for different initial amplitudes, and we report here the results for $h_0(t_0)= 2\times10^{-23}$. 
For more information, see the \textcolor{black}{Appendix} \ref{GPUcost}.

\begin{figure*}[hbtp]
    \centering
\subfloat[Number of maps]{\includegraphics[scale=0.45]{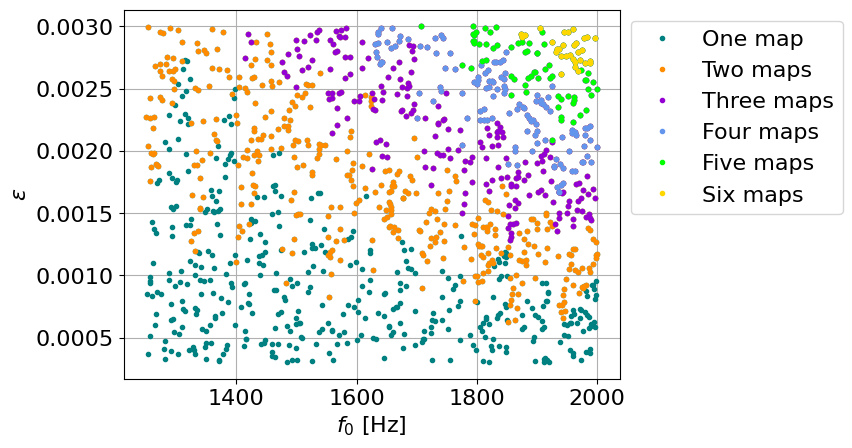}} \hfill
\subfloat[ $pr_{sn}$]{\includegraphics[scale=0.45]{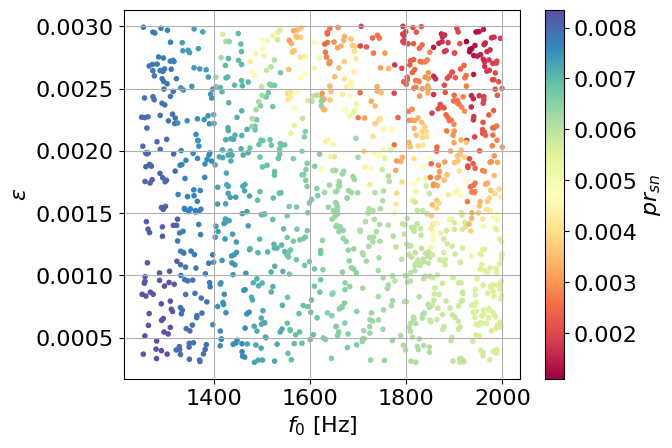}}\hfill\\
    \caption{For the denoiser training set, we plot, as a function of initial frequency and ellipticity, the number of maps crossed by the same signal (left panel), and the pixel-signal-to-noise-ratio (\textcolor{black}{right panel}). }
    \label{fig:nmappe_snr}
\end{figure*}
The two plots in \textcolor{black}{Fig.} \ref{fig:nmappe_snr} summarize the main properties of the denoiser training set.
On the left panel, we report the number of maps crossed by each signal, and on the right panel, the $pr_{sn}$ (Eq.(\ref{eq:prsn})) of each signal versus the NS ellipticity ($\epsilon $) and initial spin frequency ($f_0 $). 
Each point represents a different signal.
Notice that the higher the initial frequency and the ellipticity, the higher the number of maps crossed by the signal.~Indeed, as described in Eq. (\ref{eq:freq_variation}), the frequency in this case varies more rapidly.
As a result, the signal loses energy faster, and $pr_{\rm sn}$ is lower (see the upper right corner of the parameter space).
As discussed in the next subsection, the region where both $f_0$ and $\epsilon$ are large is where both denoising and classification become more challenging.
The test set exhibits the same trend.\par

\subsection{Denoising}\label{subsec:denoising}
We trained the denoiser as described in \textcolor{black}{Sec.} \ref{subsec:denoiser}, and then 
evaluated the model's performance in reducing the noise and in preserving the signal separately.
In particular, while the training was done with "noise plus signal" maps, the tests were carried out separately on "only noise" maps and on maps containing both noise {\it and} a signal. \par

\begin{figure*}[hbtp]
   \centering
\subfloat{\includegraphics[scale=0.5]{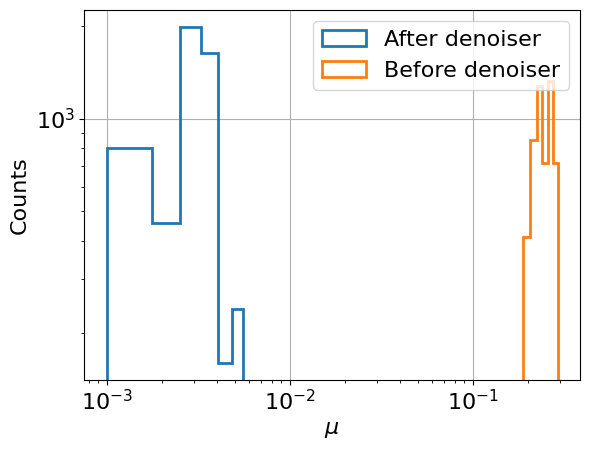}} \hfill
\subfloat{\includegraphics[scale=0.5]{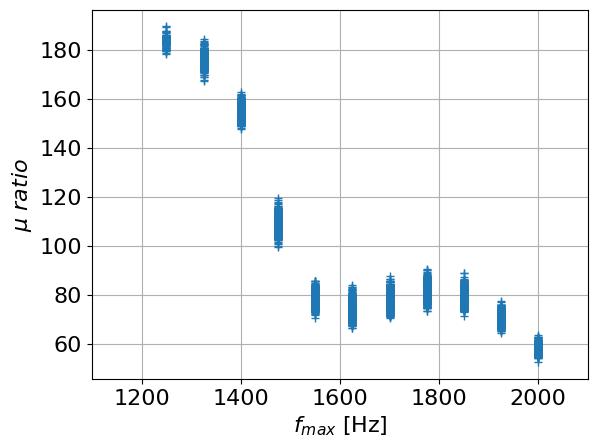}}\hfill\\
\caption{Noise reduction evaluation. On the left-hand panel, we report the histograms of the ASD pixels mean for the testing set, before (orange) and after (blue) the denoiser. On the right-hand panel, we report the ratio between the mean of the maps that have not been passed through the denoiser to the mean of the denoised maps as a function of the highest frequency of each time-frequency map.} 
   \label{fig:den_noise}
    
\end{figure*}

We first focus on the noise reduction capability and summarize our results in \textcolor{black}{Fig} \ref{fig:den_noise} .
The histograms on the left panel show the \textcolor{black}{mean value of all pixels in each map of the test set ("noise only" maps) before} ($\mu_\mathrm{noise}$, orange) and after ($\mu_\mathrm{den}$, blue) the denoiser, i.e.
\begin{equation}
\begin{aligned}
&\mu_\mathrm{noise} = \sum_{t,f}\frac{1}{N_{pix}}N_{tf}^\mathrm{noise} \ \ \ \ \ 
\mu_\mathrm{den} = \sum_{t,f}\frac{1}{N_{pix}}N_{tf}^\mathrm{den} \\
\end{aligned}
\end{equation}
where $N_{tf}$ is the same as in Eq. (\ref{eq:prsn}).
On the right panel, we show the ratio of the means, defined as 
\begin{equation}
    \mu_\mathrm{ratio}= \frac{\mu_\mathrm{noise}}{\mu_\mathrm{den}} \, 
\end{equation}
as a function of $f_{\rm max}$, i.e. the highest frequency of the time-frequency map.

We observe that after denoising, the mean values are reduced by up to two orders of magnitude and note a decreasing trend with frequency in the ratio.
The reason for the latter is the dependence of the noise ASD on frequency.
In fact, in the upper part of our frequency range, the noise ASD is higher (see, e.g., \cite{noise}), and denoising becomes more challenging. \par

\begin{figure}
    \centering
    \includegraphics[scale=0.5]{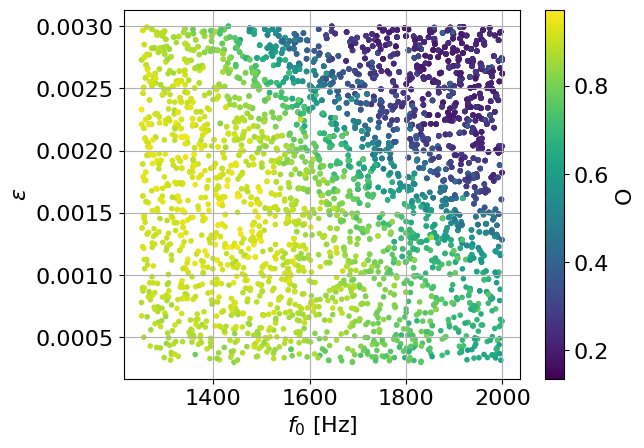}
    \caption{Best overlap values for each set of maps, as a function of the value of frequency and ellipticity.}
  \label{fig:den_over}
\end{figure}
We then turned to evaluating the effectiveness of the denoiser in preserving the signal by testing it on "noise+signal" maps. 
Given that, as discussed above, individual signals may appear in multiple maps, we report in \textcolor{black}{Fig.}  \ref{fig:den_over} the best value of the overlap, $  \mathcal{O}$ (Eq. (\ref{eq:over})), in the $f_0-\epsilon$ plane, i.e., the value of the overlap computed on the map where the signal has been denoised \textcolor{black}{the best}.
Indeed, we stress that, in order to have a valid "signal trigger," it is sufficient to be able to tag right one map.
We note that there is a correlation between the parameters of the signals and the ability of the denoiser to retrieve them.
Indeed, the worst values of overlap are found in the upper right corner, the same region where $pr_{\rm sn}$ is lowest (right panel of \textcolor{black}{Fig.} \ref{fig:nmappe_snr}), and the signal reaches the noise level more rapidly.\par
Despite this, the denoiser is capable of retrieving at least half of the signal for the majority of the parameter space.
Overall, $76\%$ of the signals have an overlap higher than 0.5.

\subsection{Classification} \label{subsec:class_res}
Our goal is to distinguish maps where the signal is present from those where it is absent.
The training procedure is applied to our classifier, described in \textcolor{black}{Sec.} \ref{subsec:class}, both on maps that were passed through the denoiser (clean maps) and maps that were not passed through the denoiser (noisy maps).
In the latter case, the training of the classifier fails (for an initial amplitude of 2E-23), i.e., the model does not find a set of parameters that minimize the training and validation losses.
As a consequence, the model does not learn the difference between maps with signals and maps that contain only noise.
This demonstrates the crucial role of the denoiser in enhancing the chances of detection. 
We thus went on to only classify clean maps, using the F1 score (Eq.~(\ref{eq:f1score})) to find the best compromise between FAP and Eff.\par 

\begin{figure*}[hbtp]
   \centering
   
\subfloat{\includegraphics[scale=0.55]{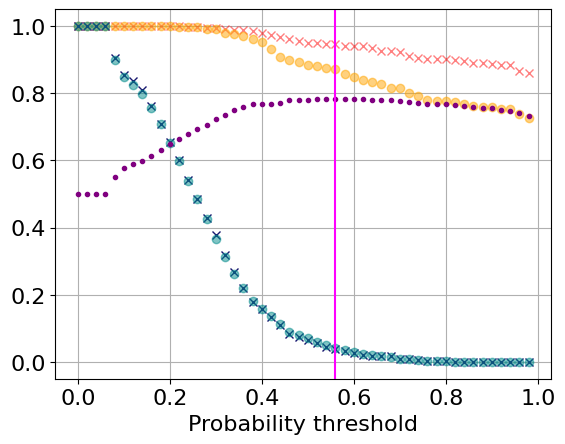}}\hfill
\subfloat{\includegraphics[scale=0.55]{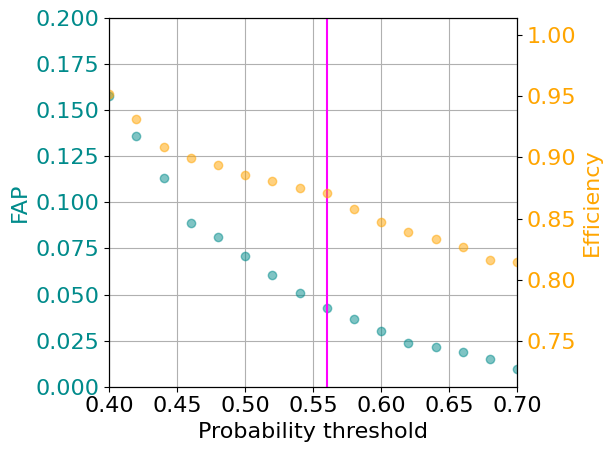}}\hfill\\
\caption{On the left panel, efficiency per signal (orange), efficiency per signal with a maximum frequency of 1800 Hz (red x), FAP (cyan), FAP with a maximum frequency of 1800 Hz (blue x) and F1 score (purple) are shown for different values of the threshold to discriminate, in each map, among the presence or absence of a signal. The straight fuchsia line corresponds to the maximum of the F1 score. On the right panel, a zoom in the range of thresholds between 0.4 and 0.7 is shown, with the FAP on the left y-axis, and the efficiency on the right y-axis.} 
   \label{fig:eff_FAP}
    
\end{figure*}

In \textcolor{black}{Fig.} \ref{fig:eff_FAP}, we show the FAP, Eff and F1 score for different values of the probability threshold of the classifier. 
The fuchsia straight line is the value that maximizes the F1 score and corresponds to 
\begin{equation}
\begin{aligned}
&  \mathrm{FAP}=4\% \ \ \ \ \mathrm{FAP}_{1800}=4\%\\
& \mathrm{Eff}=87\% \ \ \ \ \mathrm{Eff}_{1800}=94\%
\end{aligned}
\label{eq:FAP_eff}
\end{equation}
where $\mathrm{FAP}_{1800}$ is the FAP computed for noise-only maps whose frequency range is under 1800 Hz (blue crosses) and $\mathrm{Eff}_{1800}$ is the efficiency computed on signals with $f_0<1800$ (red crosses).
\textcolor{black}{As we would expect}, $\mathrm{Eff}_{1800}$ is higher than Eff and decreases slower.
Indeed, as shown in \textcolor{black}{Fig.} \ref{fig:den_over}, as the frequency increases, it is more difficult to retrieve the signal.
On the other hand, the FAP is not sensitive to the frequency and does not differ from $\mathrm{FAP}_{1800}$.

\begin{figure}[hbtp]
    \centering
    \includegraphics[scale=0.5]{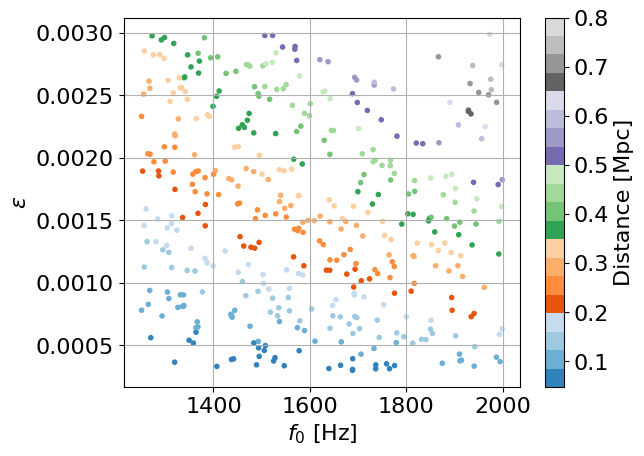}
    \caption{Distance of the source associated to the retrieved signals as a function of frequency and ellipticity.}
    \label{fig:distance}
\end{figure}

In \textcolor{black}{Fig.} \ref{fig:distance}, we show the distance of the source associated with the retrieved signals as a function of $f_0$ and $\epsilon$.
As already stated, we consider a signal retrieved when we correctly tag at least one map that it crosses.
Fewer points in the upper right corner of the maps show that these signals are more difficult to correctly label.
These signals correspond to more distant sources.
We expect this result, given the values the overlap assumes in that part of the parameter space (see \textcolor{black}{Fig.} \ref{fig:den_over}).
Anyway, we are able to classify sources distant up to 0.8 Mpc.
This is still not enough, given the LIGO O4 sensitivity curve, because according to \cite{rate_sn_Ando_2005}, at this distance, the supernova rate is lower than 0.1 per year.
In addition, we expect magnetars to account for approximately 10$\%$ of the NS population \cite{mag_population}, so the chance of having an event within these distances is low.

\subsection{Denoiser improvements: masked loss}\label{subse:masked_loss}
\begin{figure*}
\includegraphics[scale=0.37]{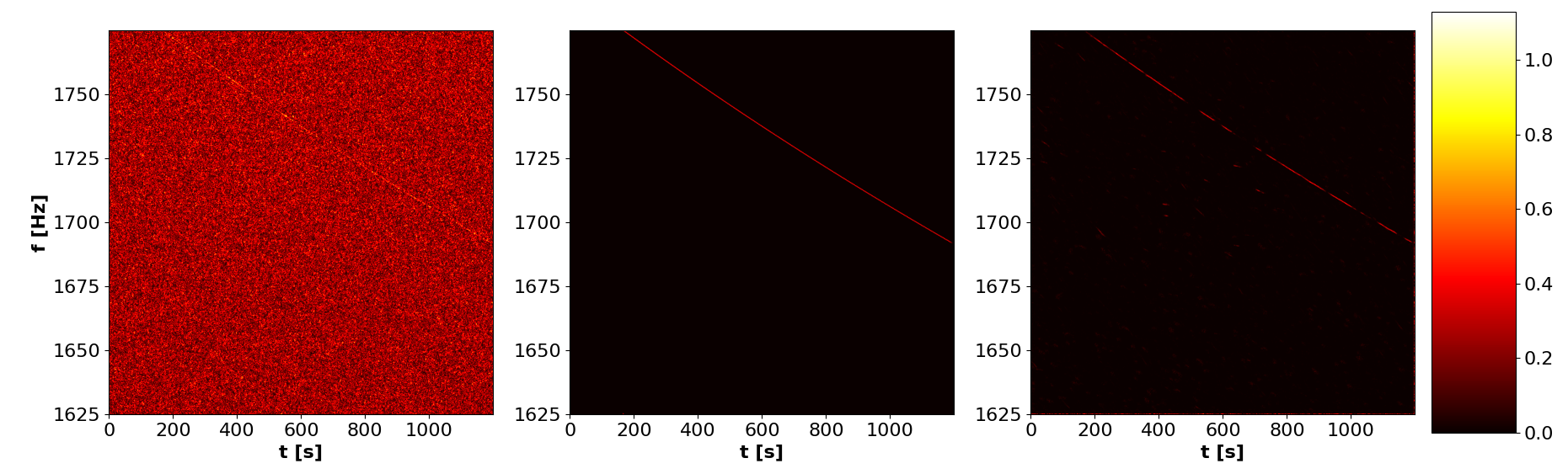}
\caption{\textcolor{black}{Example of map} containing noise plus signal, only signal and denoised noise plus signal ($\mathcal{O}=0.58$). The physical parameters of the signal are $f_0$=1791 Hz, $\epsilon$=0.001, $h_0=2\times 10^{-23}$. 
Each map has a frequency range between 1775 and 1625 Hz and covers a time interval of 1200 s. The noise is simulated according to the O4 design sensitivity curve of LIGO interferometers.}\label{fig:trittico_denoised}
\end{figure*}

We need to increase the denoiser performance to reach better FAP or Eff. 
In particular, we need to improve our ability to retrieve the signal.
During the training phase, we noticed that the loss decreased smoothly to a plateau. 
So in order to obtain better results, a change is needed, either in the NN structure or in the training procedure. 
A study of a diverse neural network pipeline will be the subject of future work. 
Here we report how we tried to highlight the structure of the signal and help the model learn it, using a masked loss.
We note that the number of pixels in a single map (360 000) is three orders of magnitude higher than the number of pixels that form the signal ($\mathcal{O}$(600)) \footnote{As we have already stated in Sec. \ref{subsec:time_freq_maps}, we chose the size of the maps taking into consideration the properties of the signal we are interested in.}. 
This suggests modifying the training scheme through the use of a mask in order to improve the learning process.
Initially, each map was multiplied by a matrix that had 1 on the signal pixels and 0 otherwise.
As the training proceeded, using a gaussian blurring function, we enlarged the area of the map that was not set to zero, increasing the difficulty of the task.
A more detailed explanation of visual attention methods, such as masked loss, can be found in \cite{attention_method}.
In \textcolor{black}{Fig.} \ref{fig:trittico_denoised}, we report an example of the output of the denoiser.
In particular, on the left-hand panel there is a map containing noise plus a signal; on the center panel there is the signal we want to retrieve; and finally, on the right hand panel there is the output of the denoiser.

\begin{figure}[hbtp]
    \centering
    \includegraphics[scale=0.5]{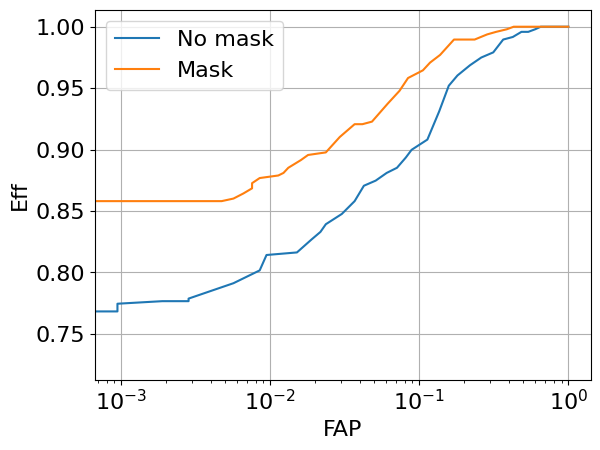}
    \caption{ROC curves to compare the classifier performance on maps that have been denoised with a masked loss (blue) and the ones without it (orange).}
    \label{fig:ROC}
\end{figure}

\begin{figure*}[hbtp]
   \centering
\subfloat{\includegraphics[scale=0.55]{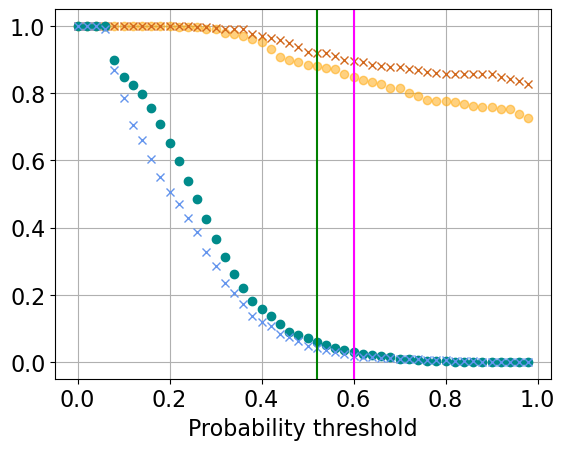}}\hfill
\subfloat{\includegraphics[scale=0.55]{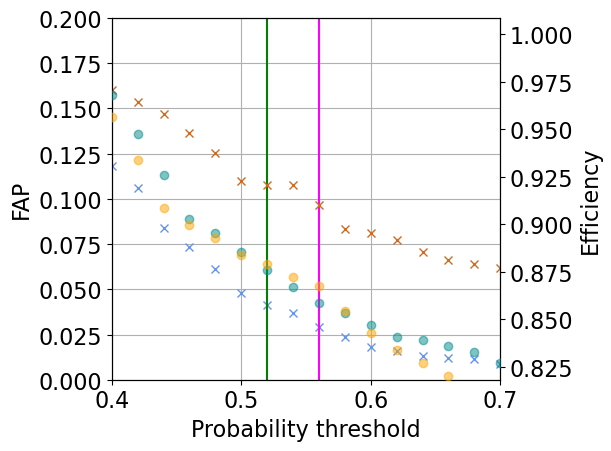}}\hfill\\
\caption{\textcolor{black}{On the left panel, efficiency per signal without masked loss }(orange, as in Fig. \ref{fig:eff_FAP}), efficiency per signal with masked loss (brown x), FAP without masked loss (cyan, as in Fig. \ref{fig:eff_FAP}) and FAP with masked loss (cornflower blue x) are shown for different values of the threshold to discriminate, in each map, among the presence or absence of a signal. The straight fuchsia line corresponds to the maximum of the F1 score. The straight green line corresponds to FAP= 4 $\%$. On the right panel, a zoom in the range of thresholds between 0.4 and 0.7 is shown, with the FAP on the left y-axis and the efficiency on the right y-axis.}
   \label{fig:eff_FAP_mask}
    
\end{figure*}

In \textcolor{black}{Fig}. \ref{fig:ROC}, we report the ROC curves, obtained by the previous training technique. 
On the x-axis, there is the FAP, while on the y-axis, there is the Eff as a function of classifier threshold.
The blue curve refers to learning without masked loss, i.e., what we have illustrated in \textcolor{black}{Sec.} \ref{subsec:class_res}.
The orange one is the best result that we have obtained after different trials with the masked loss, changing the parameters of the blurring.
In \textcolor{black}{Fig.} \ref{fig:eff_FAP_mask} \textcolor{black}{we report the FAP and Eff in the masked loss case (brown and cornflower blue x), compared with the same quantities as in } Fig. \ref{fig:eff_FAP}. We observe that the values are better:
\begin{equation}
  \mathrm{FAP}=2\% \ \ \ \ \mathrm{Eff}=90\%\\
\label{eq:FAP_eff_mask}
\end{equation}

\begin{figure}[hbtp]
    \centering
    \includegraphics[scale=0.5]{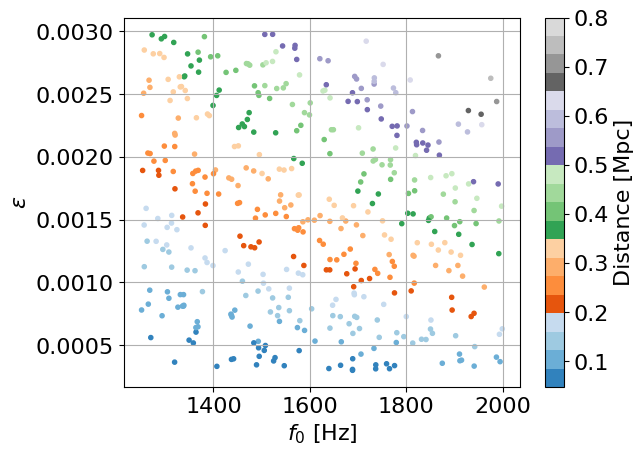}
    \caption{Distance of the source associated to the retrieved signals as a function of frequency and ellipticity, masked loss.}
    \label{fig:distance_mask}
\end{figure}

So the FAP decreases while the efficiency increases.
At the same time, in \textcolor{black}{Fig.} \ref{fig:distance_mask} we report the distance of the source associated with the retrieved signals as a function of the parameter values.
We see that even if the classifier identifies more signals, it still has problems in the upper right corner.

To compare with Eq. (\ref{eq:FAP_eff}), we can choose a FAP of $4\%$.
So adopting the masked loss training, we obtain a higher Eff, i.e.:

\begin{equation}
\begin{aligned}
\mathrm{FAP}=4\% \ \ \ \ \mathrm{Eff}=92\%\\
\end{aligned}
\label{eq:FAP_eff_mask_FAP}
\end{equation}
as shown in Fig. \ref{fig:eff_FAP_mask}, \textcolor{black}{in correspondence of the green vertical line.}
It is important to observe that we are conducting our analysis using data from a single interferometer.
By combining data from a network of detectors, we expect to be able to lower the FAP.
In fact, if we train our models separately on different interferometers data and then combine the results, we should lower our FAP down to the product of the different FAPs in the best-case scenario.
This would allow us to choose a less stringent threshold to classify a map as having a signal.

\subsection{Changing the braking index}\label{subsec:braking_index} 
\textcolor{black}{At the beginning of our work we trained our model assuming a braking index $n=5$ (see equation (\ref{Eq:spin_down_general})), that is, without taking into account the electromagnetic contribution to the spin-down (which is described, for instance, in \cite{Dall_Osso_braking_index}).
To assess the robustness of the model, we tested its performance by injecting signals with \textcolor{black}{different} values of $n$, since
our goal in the future is to generalize our technique by combining electromagnetic and gravitational torques.}\par
\begin{figure}
    \centering
    \includegraphics[scale=0.5]{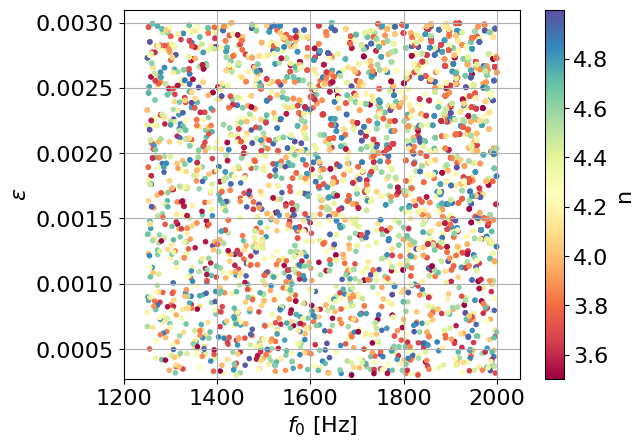}
    \caption{\textcolor{black}{Parameter space} of the signals we have simulated, on the y-axis there is $\epsilon$ and on the x-axis $f_{0}$, while on the colorbar there is $n$.}
    \label{fig:brake_parameters}
\end{figure}
\begin{figure*}[hbtp]
   \centering
\subfloat{\includegraphics[scale=0.55]{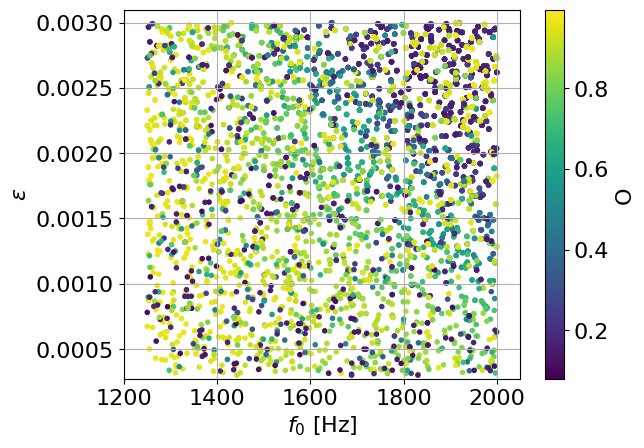}}
\subfloat{\includegraphics[scale=0.55]{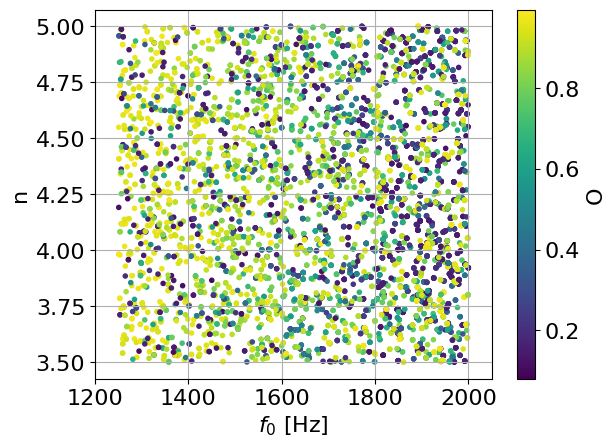}}\\
\caption{Best overlap values, on the left-hand panel, as a function of the value of $f_{0}$ and $\epsilon$ and on the right-hand panel , as a function of $f_{0}$ and $n$.}
   \label{fig:over_nbrake}
\end{figure*}

\textcolor{black}{So using the models already trained, we tried to test the denoiser and later train and test the classifier on signals, with the same} $f_0$ and $\epsilon$, with n randomly taken in the range [3.5, 5].
It is important to be able to retrieve signals with different braking indexes in order to take into account the different emission processes of the star.
In particular, we would like to have the capability of finding signals in regimes that differ from purely gravitational emissions.
We did not train the denoiser again; rather we have used the best model of the previous subsection, trained with $n$ = 5.
Then, with the new denoised maps, we trained the classifier again.
The idea is to show that even if we make a restrictive assumption, the model is able to retrieve different power laws from the one we trained it with.
In \textcolor{black}{Fig.} \ref{fig:brake_parameters}, we show the new parameter space, where $f_0$ and $\epsilon$ are the same as the previous section, while we changed $n$.
\textcolor{black}{For what it concerns the fraction of the pixels of the signal which is} preserved, in \textcolor{black}{Fig.} \ref{fig:over_nbrake}, we report the best value of the overlap for each set of maps as a function of initial frequency, ellipticity and braking index.

From the right-hand panel, we can notice that there is no trend in the braking index, in fact, the values of the overlap do not seem to depend on $n$.
Nevertheless, on the left-hand panel, we notice the usual trend in $f_{0}$ and $\epsilon$. 
However, the different braking index makes the classification procedure easier in the upper right corner of the parameter space while worsening the situation in the middle, as we can notice comparing with \textcolor{black}{Fig.} \ref{fig:den_over}.\par
\begin{figure*}[hbtp]
   \centering
\subfloat{\includegraphics[scale=0.55]{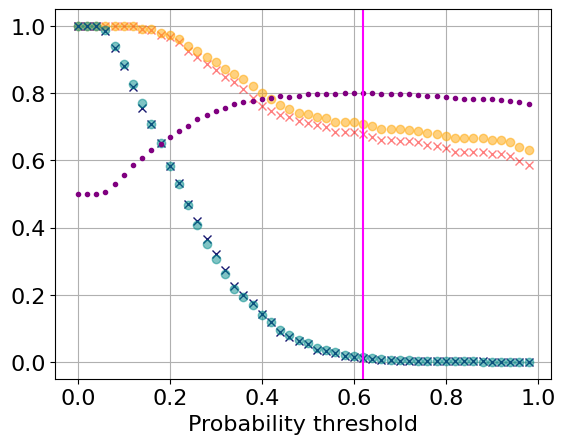}}\hfill
\subfloat{\includegraphics[scale=0.55]{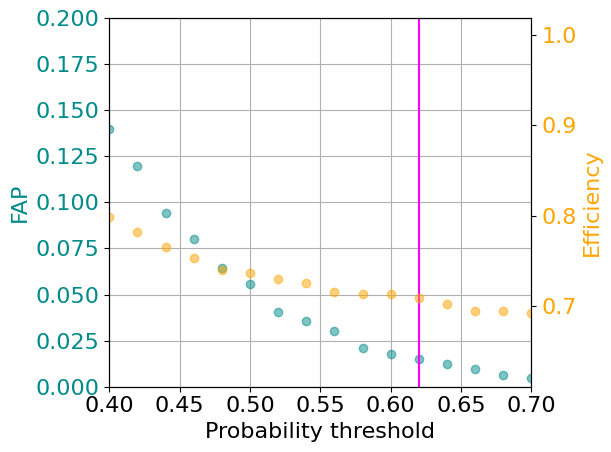}}\hfill\\
\caption{Same quantities as in Fig. \ref{fig:eff_FAP}. \textcolor{black}{The denoiser has been trained with the masked loss on signal with} $n$ = 5. The classified signals have n $\in [3.5,5]$.}
   \label{fig:eff_FAP_nbrake}
\end{figure*}

As a consequence, the classification task is more difficult.
In \textcolor{black}{Fig.} \ref{fig:eff_FAP_nbrake}, we report the values of efficiency, FAP and F1 score for different thresholds; the maximum of the F1 score (fuchsia line) corresponds to
\begin{equation}
\begin{aligned}
&  \mathrm{FAP}=1\% \ \ \ \ \mathrm{FAP}_{1800}=1\%\\
&\mathrm{Eff}=71\% \ \ \ \ \mathrm{Eff}_{1800}=67\%.
\end{aligned}
 \label{eq:FAP_eff_nbrake}
\end{equation}
in this case, the efficiency with a threshold on the initial frequency is slightly worse.

The efficiency is worse than (\ref{eq:FAP_eff}, \ref{eq:FAP_eff_mask}), but we gain a better FAP, so fixing the FAP at $4\%$, as in the first case Eq. (\ref{eq:FAP_eff}), we obtain:
\begin{equation}
\begin{aligned}
&  \mathrm{FAP}=4\% \ \ \ \ \mathrm{FAP}_{1800}=4\%\\
&\mathrm{Eff}=73\% \ \ \ \ \mathrm{Eff}_{1800}=70\%.
\end{aligned}
 \label{eq:FAP_eff_nbrake_FAP}
 \end{equation}
As in the previous case, Eq. (\ref{eq:FAP_eff_mask_FAP}), the efficiency increased.\par
We demonstrated that the method presented in this paper is robust to different values of $n$, ie. different power laws, even if we trained the denoiser with a fixed braking index. 

\subsection{Computational load}
The total computing time for all the tasks described in this paper has been of the order of several hours.
In particular, the creation of all the maps ($\mathcal{O}$(7000)) took about 6 hours.
In addition, we needed 3 hours to train the denoiser with $\mathcal{O}$(2000) maps and 20 minutes to denoise $\mathcal{O}$(5000) maps.
To train the classifier, 30 minutes were needed on $\mathcal{O}$(4000) maps, and the classification task took less than 5 minutes on $\mathcal{O}$(1000) maps.
In other words, to train both models, we needed $\mathcal{O}$(6000) maps of 1200 s that correspond up to 40 days of data.
For more information, see \textcolor{black}{Appendix} \ref{GPUcost}.

\subsection{Discussion}
\textcolor{black}{In \cite{gw170817} a search for a tCW from a possible newly formed magnetar was conducted following the discovery of GW170817.
In that case, different parameters were used for the simulation of the signal, in particular a moment of inertia that was three times bigger than the value we have used (at the limit of the mass ranges described in literature).
As it is possible to notice in Eq. }(\ref{eq:wave_amplitude_ellipsoid}),\textcolor{black}{ the higher the moment of inertia, the higher the amplitude, and the easier it is to classify signals}.\par
\textcolor{black}{A fair comparison of our method to the methods employed in \cite{gw170817} is not straightforward because the considered parameter space is different in the two cases, and there are also differences in the assumed NS moment of inertia. 
However, we ran a code to estimate the optimal sensitivity of the }\textit{generalized Frequency Hough (GFH)} \textcolor{black}{pipeline (see \cite{gfh_method}, Eq. (34)).
We used the same signals, under the same simulated $O4$ noise. 
We obtain a maximum distance that is slightly less than twice that obtained by our method.
It should be kept in mind, that this result is obtained under ideal conditions, in which the entire signal is fully integrated into the GFH calculation and the optimal data segment duration is used in all the parameter space.
This approach is not realistically applicable for a general analysis in which signal parameters are not known a priori.
A systematic comparison of these two methods sensitivities is beyond the scope of this article.
Overall, we are confident that our technique will allow us to improve future searches to identify tCW candidates.}
\par

\section{Conclusions}\label{sec:conclusions}

The goal of this study was to develop machine learning techniques for the search of tCW signals, such as those emitted by newborn magnetars, in interferometric data.
We built, using a machine learning approach, a classifier to split time-frequency maps into two categories: presence or absence of signals.
To help with the classification task, we built a denoiser.\par
To test the performance of our method, we simulated time-frequency maps containing noise plus signals.
The noise was simulated according to the LIGO O4 design sensitivity curve, as explained in \cite{colored_noise}, and the signal according to the rotating ellipsoid model (e.g. \cite{Maggiore}) with $n=5$.\par
Here we report our results for signals having an initial amplitude of $2 \times 10^{-23}$, the minimum value at which the use of the denoiser was beneficial to train the classifier. \par
We illustrate how this procedure can be used to search for tCWs emitted by newly born magnetars, and show that the denoiser plays a crucial role in the successful operation of the classifier.
Indeed, we are not able to classify maps that have not been passed through the denoiser at an initial amplitude of $2 \times 10^{-23}$.
Moreover, the procedure proved to be robust to different power laws with $n\in[3.5,5]$, even if the denoiser was trained with $n$ = 5. 

    

\textcolor{black}{An extremely relevant point is the rapidity of this technique and the limited requirements in terms of computing power.
In fact, we are able to analyze the equivalent} of $\mathcal{O}$(14) days of data in a few minutes.\par
In addition, 40 days of interferometer data should be sufficient to train the whole NN pipeline, and it is possible to use the rest of a scientific run to make inferences.
Our method needs a number of maps to train the model that is about one order of magnitude smaller than in \cite{miller}.\par

\textcolor{black}{Overall, the proposed method has a sensitivity which is comparable with the already existing methods.
In general, more sensitive and model-dependent methods can be devised to perform a follow-up of signal candidates and measure signal parameters.}\par

In the future, we plan to do some further studies. In particular, we would like to find an efficient approach to combining the data of different interferometers and conduct a deeper study on cases with a different braking index.
At the same time, we aim at improving our denoiser by trying different NN architectures.
Our goal, by combining these improvements with the new generation of detectors and the new observing run, is to be able to reach distances at which the probability of seeing this kind of event is significant.
\begin{acknowledgments}
We thank our Sapienza Physics Department colleagues, S. Giagu and A. Ciardiello, for the useful discussion and suggestions.
We also thank INFN and the Amaldi Research Center for the clusters hosted in the INFN Rome infrastructure, where we have stored the data used in this work and run the present analysis. We thank the INFN-CNAF computing staff for the resources we have used in this analysis and for their constant support.~SD acknowledges funding by the European Union’s~Horizon2020 research and innovation programme under the Marie Skłodowska-Curie (grant agreement No.754496).
\end{acknowledgments}

\appendix

\section{Denoiser architecture}
\label{denoiserArch}
\begin{itemize}
    \item First layer: 64 convolutional filters of size 3x3x1 and ReLU as activation function, i.e. 1 input channel and 64 output channels.
    \item Second group of layers: six layers, each one with 64 convolutional filters of size 3x3x64, i.e. 64 input channels and 64 output channels. Later, we have batch normalization and finally ReLU as an activation function.
    \item Last layer: 1 convolutional filter of size 3x3x64, i.e. 64 input channels and one output channel.
\end{itemize}
\section{Classifier architecture}
\label{classifierArch}
\begin{itemize}
\item First layer: 5 convolutional filters of size 10x10x1 and ReLU as activation function, i.e. 1 input channel and 5 output channels. The filters move with a stride of 5.
\item Second layer: 10 convolutional filters of size 6x6x1 and ReLU as activation function, i.e. 5 input channels and 10 output channels. The filters move with a stride of 3.
\item Third layer: 1 max pooling layer with a kernel of size 5x5x1.
\item Fourth layer: one linear layer that reduces the dimension from 1690 to 84 and ReLU as an activation function.
\item Last layer: a linear layer that passes from 84 numbers to 2 and softmax as activation function.
\end{itemize}
\section{GPU usage and NN training}
\label{GPUcost}

To speedup the NN training and inference times, we used an NVIDIA L40S GPU with 48GB of RAM, installed on the Virgo INFN Rome cluster.
The denoiser for the first train, \textcolor{black}{Sec.} \ref{subsec:denoising}, was trained for fifty epochs, with a batch size of 8. 
Beyond that, the loss did not show significant improvement.
After the implementation of the masked loss, \textcolor{black}{Sec.} \ref{subse:masked_loss}, we trained the model for 200 epochs.
In the first 160, we used masked loss, changing the width of the blurring function every 16 epochs. 
In the last 40 epochs, we trained it using the whole map while computing the loss function.
As in the previous case, we used as batch size 8.\par
\textcolor{black}{Concerning} the classifier, we trained for 40 epochs with a batch size of 40 in the first two scenarios, Secs. \ref{subsec:class_res},\ref{subse:masked_loss}, and 60 epochs in the last one,\textcolor{black}{Sec.} \ref{subsec:braking_index}. 

\newpage
\bibliography{main}

\end{document}